\documentclass{appolb}

\usepackage{graphicx}
\usepackage{wrapfig}

\begin{document}

\title{Optimizing Polar Angle Asymmetry Observables at Colliders
\thanks{Presented by SD at the Light Cone 2012 conference, Cracow, Poland, 8-13 July}
}
\author{S. Dalley, S. Adhikari,  P. Nadolsky
\address{Department of Physics, Southern Methodist University
\\
Dallas, 
TX 75275-0175, U.S.A}
}
\maketitle
\begin{abstract}
Angular asymmetries are simple, intuitive, model-independent 
observables used to identify spins of new elementary particles. 
In the case of Drell-Yan-like boson resonances, 
we generalize the well-known center-edge angular asymmetry to optimize
spin identification when only a limited sample of events is available. 
By choosing simple weight functions $W(\theta)$ in integrals over the 
polar angle $\theta$, such as $W = \cos^n \theta$, 
we can improve spin discrimination significantly in production and decays 
of spin 0, 1, and 2 bosons.  The power $n$ can be tuned 
in particular cases, but $n=2$ ($n=1$) works 
well for any forward-backward symmetric (non-symmetric) decay to massless 
particles.  
\end{abstract}



New particles, such as Higgs, $Z^\prime$, excited gravitons, sparticles, etc., 
may well be discovered in collider data from 
$p \bar{p}$ (Tevatron), $pp$ (LHC), or $e^+ e^-$ (ILC) scattering.
The spin of a new virtual particle is often determined in the experiment 
from the analysis of angular distributions in the heavy particle's decay. 
This analysis follows a standard quantum-mechanical expansion 
over spherical harmonics that depends on the spin of the massive particle 
state. In this contribution, we describe an analysis procedure 
for decay angular distributions that is more efficient 
in using limited event statistics than the commonly adopted method.  
One way to determine the spin is to exploit the ``center-edge'' 
asymmetry that typically exists between forward and 
transverse 
decay products from a boson \cite{CE1,CE2}. We show how to modify 
the usual ``central-edge'' asymmetry definition to increase its discriminating power for the spin. We keep the discussion transparent by working at the parton 
level, but consider a simple analytic model for typical experimental 
acceptance. We checked that our main conclusions remain valid in a more realistic calculation, by comparing our method to a fully differential calculation 
using ResBos \cite{resbos} that includes proton PDFs, NLO corrections, 
and NNLL resummations in QCD.

For decay of a boson into two back-to-back massless particles, 
we define a general asymmetry by an integral
\begin{equation}
A = \int_{-1}^{1} W(z)\, P(z)\, dz.
\label{asym}
\end{equation}
Here $z=\cos \theta$, $\theta$ is the decay polar angle in the boson 
center-of-mass frame, 
$P(z) = \frac{1}{\sigma} \frac{d \sigma }{ d z}$ is the normalized 
production probability density, and $W(z)$ is a weight function. One popular
choice for $W(z)$ has been 
the center-edge step function, \cite{CE1, CE2, CE3, CE4, CE5, CE6,rapid1}  
\begin{equation}
W_{\rm CE}(z) = \  +1 \ \ (|z| > z^*); \  -1  \ \ (|z| < z^*),
\label{step}
\end{equation}
with $z^* \sim 0.5$. We will propose 
an alternative definition for $W(z)$ that is more optimal than $W_{\rm CE}(z)$.

To discriminate between two possible boson spins $a$ and $b$ in a measurement,
we introduce respective probability densities $P_a(z)$ and $P_b(z)$ and their 
asymmetries $A_a$ and $A_b$, and construct the ratio 
\begin{equation}
R_{ab} = \frac{A_b - A_a}{\delta A_a + \delta A_b}.
\label{argh}
\end{equation}
The expected statistical uncertainty is estimated  by 
\begin{equation}
\delta A = \sqrt{\int P(z)\,W(z)^2\,dz - \left(\int P(z)\,W(z)\, dz\right)^2}.
\label{stat}
\end{equation}
$R_{ab}$ is independent of additive and multiplicative constants in $W(z)$. 
A larger $R_{ab}$ value indicates that $A$ with the chosen weight 
is more sensitive to spin. We wish to choose $W(z)$ to maximize $R_{ab}$, 
but we don't want to fine-tune $W(z)$ 
for each particular process/model/experiment. 
We will show for typical cross-sections of interest that the simple 
choice $W(z) = z^n$ works well. Although $n$ may be tuned, $n=2$ ($n=1$)
is largely sufficient for distributions that are forward-backward (non) 
symmetric under $z \rightarrow -z$.

A choice of $W(z)$ that amplifies in the range of $z$ where 
the difference $\Delta  P(z) = P_b - P_a$ is large in magnitude 
can increase the numerator in eq.~(\ref{argh}).
(Note that $\Delta P$ is normalized to zero, so it must change sign).
However, amplifying this range of $z$ will unbalance 
the cancellation of statistical fluctuations 
across the whole $-1\leq z\leq 1$ range and increase the denominator
in eq.~(\ref{argh}), which must happen since $\Delta P$ is normalized. 
On the other hand, a uniform choice, such as $W_{\rm CE}$ in eq.(\ref{step}), 
gives equal weight to the whole $z$ range, reducing the denominator
through cancellation of fluctuations but not optimimizing the numerator.
It appears there is a minimization problem to solve. 

\begin{wrapfigure}{r}{0.55\columnwidth}
{\includegraphics[width=2.5in]{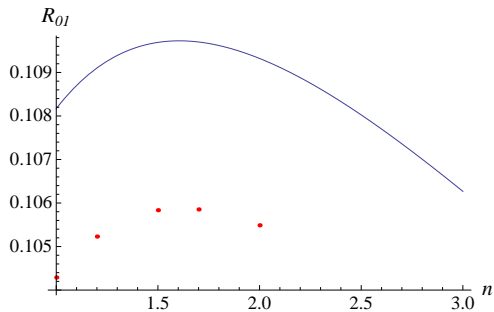}}
\caption{
The solid line is the statistical significance ratio $R_{01}$ for the 
particular case of spin 0 and  1 bosons decaying to forward-backward symmetric 
fermions. $R_{01}$ is plotted as a function of $n$ in the weight functions 
$W = |z|^n$. The points are obtained for discrete choices 
of $n$ after modulating the leading-order densities $P_{a,b}$ by
a function (\ref{accept}) simulating experimental acceptance, for
$\alpha = 6$ and $\beta = 0.2$.}
\label{Maximum}
\end{wrapfigure}
Let us consider a concrete example. We compare the lowest order QCD
densities $P_{0}$ and $P_{1}$ for spin-0 and spin-1 
bosons produced by massless partons and decaying to massless
fermion pairs:
\begin{eqnarray}
&& P_0  = \frac{1}{2}, \nonumber\\ 
&&P_1  =  \frac{3}{8} (1 + z^2+ 2\frac{c -d}{c +d}z).\nonumber
\end{eqnarray}
The term $\frac{c-d}{c+d}$ arises for general
boson-fermion chiral couplings, {\it e.g.}, it is nonzero 
in a parity-violating decay. 
Consider first the symmetric case $c = d$, so that $P_1=(3/8)\cdot(1+z^2)$. 
We try a class of weight functions\begin{equation}
W(z) =  |z|^n,
\label{trial}
\end{equation}
with $n$ selected to emphasize contributions from the $z$ intervals with
large $\Delta P(z)$. Figure~\ref{Maximum} confirms the existence of an optimal 
weight that maximizes $R_{01}$. Recalling that confidence limits are determined 
by $R\sqrt{N}$ in the case of $N$ sample events, $W_{\rm CE}$ would require 
about 1/3 more events to achieve the same level of significance as the optimal 
weight -- 
see Table~\ref{table1}. 
A similar conclusion holds for parity violating processes ($c\neq d$), with
the results for the maximal symmetry violation $d = 0$ 
shown in Table~\ref{table1}. [In this case, 
we modified the center-edge weights as 
$W(|z|) \rightarrow {\rm sign}(z)\cdot  W(|z|)$].
Provided there is a significant monotonic variation of 
$\Delta P$ from the center to the edge of the range of $z$, 
the optimum weight will be close to quadratic
 for $c = d$ and linear for $d = 0$.

\begin{table}
\begin{tabular}{|c|c|c|c|c|}
\hline 
\multicolumn{2}{|c|}{$B\rightarrow f\overline{f}$} &$R_{ab}$ with $W_{CE}$& Optimal $n$, $R_{ab}(n)$ & \% more events with $W_{\rm 
CE}$
\tabularnewline
\hline 
$a$ & $b$ & \multicolumn{1}{l|}{} &  & \tabularnewline
\hline
\multicolumn{1}{|c|}{0} & 1 & 0.095 & 1.6, 0.1097 & 34\%\tabularnewline
\hline 
0 & 2{*} & 0.106 & 4, 0.155 & 113\%\tabularnewline
\hline 
1 & 2{*} & 0.205 & 3, 0.264 & 66\%\tabularnewline
\hline 
0 & 1{*}{*} & 0.45 & 0.8, 0.52 & 33\%\tabularnewline
\hline 
\multicolumn{2}{|c|}{$B\rightarrow \gamma$$\gamma$} &  &  & \tabularnewline
\hline 
0 & 2{*} & 0.239 & 1.3, 0.272 & 30\%\tabularnewline
\hline 
\multicolumn{5}{|c|}{{*}({$\epsilon_q=0.1$}) 
and {*}{*}($d = 0$). }\tabularnewline
\hline
\end{tabular}
\caption{
The $R_{ab}$ ratio for a boson $B$ decaying 
to difermions $ff$ or diphotons $\gamma \gamma$, computed for various
spins $a$ and $b$, using $W_{CE}$ and $W=|z|^n$ weights with the
optimal $n$ value. Also shown is the percentage increase in the number
of events that would be 
needed when using $W_{\rm CE}$ to achieve the same 
statistical significance as with the optimal $W=|z|^n$.}
\label{table1}
\end{table}

\begin{figure*}
\hspace{0.9in}
Spin 0
\hspace{2.2in}
Spin 1
\\

\raisebox{-0.5in}
{\includegraphics[width=5in]{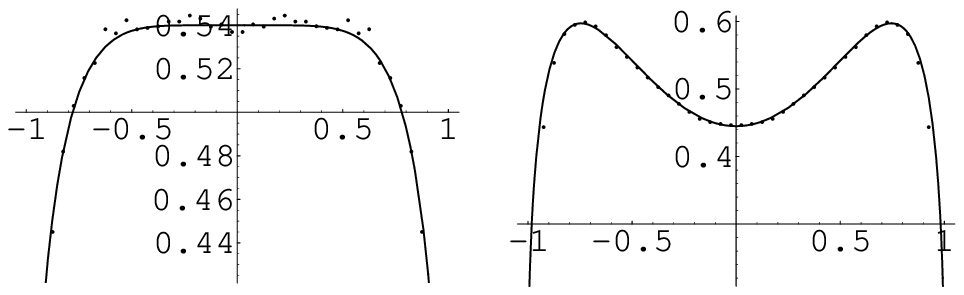}}
\caption{
Realistic $P(z)$ for spin-0 (left graph) and spin-1 (right graphs)
bosons decaying to massless fermions. The data points were 
generated using ResBos \cite{resbos} for boson mass $M= 1$ TeV produced in $pp$
collisions at $\sqrt{s} = 14$ TeV. This includes NLO and NLL QCD corrections, 
CTEQ 6.6 PDFs \cite{Nadolsky:2008zw}, and a $p_T > 25$ GeV cut on the fermion's 
transverse momentum. The solid lines represent a fit 
using a modulated $P(z)$ with $m(z)$ from eq.~(\ref{accept}).}
\label{rb}
\end{figure*}

These conclusions stay valid when the above 
lowest-order numerical estimates for $R_{ab}$ are
modified by higher-order QCD corrections to $P(z)$,
smearing by parton distributions in the case of hadronic
collisions, and detector acceptance constraints.
For example, the angular dependence of $P(z)$ can be modified by the
limits $y_{\rm min} < y < y_{\rm max}(z)$ on the boson rapidity.
We can explore the impact of these corrections by multiplicatively
modulating the lowest-order parton-level
$z$-dependence of the differential cross section by a function 
\begin{eqnarray}
m(z) =   \int_{M - \Delta M}^{M + \Delta M} dM 
\int_{-\infty}^{\infty} dy  \ \Theta(y_{\rm max} - y)\times
\Theta(y - y_{\rm min}) K(y, M)
\end{eqnarray}
with a boson mass $M$, $\Delta M$ bin size around the resonance peak,
step functions $\Theta(y)$ indicating acceptance constraints imposed on
the rapidity, and a function $K(y,M)$ containing all details of QCD corrections,
parton distributions, boson propagator, etc. For a typical $z$ distribution,
$m(z)$ can be modelled by
\begin{equation} 
m(z) \propto (1-z^\alpha)^{\beta},
\label{accept}
\end{equation}
where $\alpha$ and $\beta$ depend on details of the calculation. 
Fig.~\ref{rb}
compares representative angular distributions for fermionic decays 
of massive bosons obtained with ResBos 
for a particular parameter set, which realistically includes the effect of
PDFs, QCD NLO and NLL corrections, and acceptance cuts, against calculations
using the modulation model (\ref{accept}). 
We see that the $m(z)$ function with $\alpha$ and $\beta$ 
fitted to the ResBos points  captures 
the general features of the ResBos predictions.
For our purposes 
of comparing the statistical performance of different polar weights,
the precise values of the powers in (\ref{accept})
are not found to be crucial.
The effect  of this modulation on the polar asymmetry is 
also shown in Fig.~\ref{Maximum}. 
Although the statistical significance
is reduced generally, as might be expected because the acceptance reduces
the center-edge asymmetry, 
the conclusions about the optimum weight are essentially unchanged.

To demonstrate limitations of the above conclusions in 
relation to the overall amount of  
center-edge symmetry,  
we may look at the decay distribution of a spin-2 boson to fermions,
\begin{equation}
P_2 = \frac{5}{8}(1-3z^2 + 4z^4) \epsilon_q + \frac{5}{8} (1-z^4) \epsilon_g.
\label{P2l}
\end{equation}   
This depends in addition on the fractions of Drell-Yan events $\epsilon_q$
and $\epsilon_g$ produced via $q \bar{q}$ and gluon-gluon fusion respectively.
Under the constraint $\epsilon_q + \epsilon_g = 1$
at a $pp$ collider, $\epsilon_q$ typically varies from $\epsilon_q \sim 0$
at very low boson masses $\ll 1$ TeV 
to  $\epsilon_q \sim \frac{1}{2}$ at  4 TeV. 
We find that the statistical 
significance of an optimized power law $W=|z|^n$ is generally 
much better than $W_{\rm CE}$ 
across this range (see for example Table~\ref{table1}). 
However, the advantage
reduces as $\epsilon_q$ grows. This is because at large 
$\epsilon_q$ 
there is no longer any center-edge asymmetry at all 
and center-edge type observables are 
generally less useful. The optimal power is of course dependent on 
$\epsilon_q$, but a simple universal quadratic choice $W = z^2$ works 
very nearly as well. 

Another important decay mode for identifying boson resonances is to two prompt
photons, which are relatively clean to identify. The corresponding tree level
probability densities in this case are
\begin{eqnarray}
P_0  & = & \frac{1}{2}, \ \ \  P_1  =  0, \ \ P_2  = \frac{5}{32}(1+6z^2 + z^4) \epsilon_g + 
\frac{5}{8} (1-z^4) \epsilon_q. 
\end{eqnarray}
The statistical significance of the
optimized power law and $W_{\rm CE}$ choices for spin 2 vs. spin 0 
with acceptance modeled by 
(\ref{accept}) follows a pattern similar to the di-fermion final state (see
Table~\ref{table1}). 

Aside from the simple powers, we have also investigated other possible angular 
weights, particularly for those that strongly enhance the regions where 
$\Delta P$ is large, and find they are not superior 
in statistical significance. Choosing a weight function 
that was orthogonal to one of the two raw probability distributions used in any 
comparison did not produce better results for the $R_{ab}$ ratio when the 
experimental acceptance was taken into account; part of the reason is that 
the latter destroys orthogonality. To conclude, the sensitivity of the
$z=\cos\theta$ distribution to the spin of the decaying heavy boson 
is increased by generalizing the center-edge asymmetry 
via eq.~(\ref{asym}) with the weight function $W(z) = z^n$ and $n=2$ for 
a forward-backward symmetric ($n=1$ for a forward-backward
non-symmetric) distribution. These findings are best noticed in a
leading-order analysis of the decay angle dependence, but 
they are robust against radiative and acceptance corrections arising in
the full NLO differential calculation.

\vspace{1\baselineskip}

\noindent 
{\bf Acknowledgements:} 
SD thanks Profs. C. S. Lim and H. Sonoda, Kobe University, for hospitality
during a part of this work. This work was supported by the U. S. 
Department of Energy under grant DE-FG02-04ER41299, by the U.S. DOE
Early Career Research Award DE-SC0003870, and by Lightner-Sams
Foundation. SA was supported by an SMU Undergraduate Research
Assistantship and a Hamilton Scholarship.

\end{document}